\documentclass[showdate]{jpsj-suppl}
\usepackage{times}
\usepackage{color}
\usepackage[latin9]{inputenc}
\usepackage{amsmath}
\usepackage{amssymb}
\usepackage{epsfig}

\newcommand{\avg}[1]{\left< #1 \right>} 

\title{Raman scattering as a probe of charge nematic fluctuations in iron based superconductors}
\recdate{September 30, 2013}
\author{Y.-X. Yang$^{1}$, Y. Gallais$^{1}$\thanks{E-mail address: yann.gallais@univ-paris-diderot.fr}, R. M Fernandes$^{2}$, I. Paul$^{1}$, L. Chauvière$^{1}$,
 M.-A. Méasson$^{1}$, M. Cazayous$^{1}$, A. Sacuto$^{1}$,
D. Colson$^{3}$ and A. Forget$^{3}$} 

\inst{$^{1}$Laboratoire Matériaux et Phénom$\grave{e}$nes Quantiques
(UMR 7162 CNRS), Université Paris Diderot-Paris 7, Bat. Condorcet,
75205 Paris Cedex 13, France,\\
 $^{2}$School of Physics and Astronomy, University of Minnesota,
Minneapolis, 55455, USA\\
 $^{3}$CEA-Saclay, IRAMIS, Service de Physique de l'Etat Condensé
(SPEC URA CNRS 2464), F-91191 Gif-sur-Yvette, France 
}

\abst{We report Raman scattering measurement of charge nematic fluctuations in the tetragonal phase of BaFe$_2$As$_2$ and Sr(Fe$_{1-x}$Co$_x$)$_2$As$_2$ (x=0.04) single crystals. In both systems, the observed nematic fluctuations are found to exhibit divergent Curie-Weiss like behavior with very similar characteristic temperature scales, indicating a universal tendency towards charge nematic order in 122 iron-based superconductors.}

\begin{document}
\maketitle

\section{Introduction}
In a nematic phase, the rotational invariance is broken while the translational symmetry remains intact. Widely studied in classical soft condensed matter systems like liquid crystals \cite{Prost}, the existence of their quantum analogue, electron nematics, has also been postulated in various strongly correlated systems \cite{Fradkin}. Examples include quantum Hall systems, underdoped cuprates, bi-layer ruthenate and heavy fermions \cite{Lilly,Daou,Borzi,Okazaki}. In these systems, rather than a breaking of the continuous rotational symmetry as in liquid crystals, electronic nematic order usually implies a breaking of the discrete C$_4$ tetragonal symmetry.
More recently, the strongly anisotropic electronic properties of iron-based high temperature superconductors have been linked to the existence of a underlying electron nematic order  \cite{Chu,Kasahara}. In AFe$_2$As$_2$ (A=Ba,Sr,Ca) (122 family) compounds, a structural transition in which the Fe-Fe bonds in the $x$ and $y$ direction become inequivalent occurs simultaneously, or precedes, a magnetic transition below which  the system displays stripe-like magnetic ordering with antiferromagnetic (ferromagnetic) alignment along the longer (shorter) Fe-Fe bonds. The structural transition between a tetragonal and an orthorhombic phase, observed in close proximity to superconductivity, is believed to be a direct consequence of electron nematic ordering \cite{Fang,Xu,Fernandes-shear}.
\par
Evidences that the structural transition is a consequence of electron nematic ordering come mostly from measurements under uni-axial stress in the orthorhombic phase. Transport, ARPES and optical conductivity measurements on mechanically detwinned crystals have revealed strong in-plane electronic anisotropies which cannot be ascribed to the small orthorhombic distortion alone \cite{Chu,Tanatar,Chu-2,Dusza,Nakajima-PNAS,ARPES-Yi}. However since these experiments were performed under a symmetry breaking field, uni-axial stress, which already breaks the $C_4$ symmetry, most of them could not unambiguously determine whether the structural transition is a true nematic phase transition, nor the exact nature of the primary nematic order parameter, spin or charge/orbital \cite{Fernandes_SUST,Paul, Kontani,Lee}.  In the case of transport, the observed anisotropies seem to depend strongly on disorder, raising the question of the role played by scattering mechanisms \cite{Ishida}. 

\par
Another route for studying nematicity is to probe the associated response functions in the lattice, spin or charge sector in the tetragonal phase in the absence of any symmetry breaking field. In the case of the lattice, orthorhombic softening has been observed for a wide range of doping via shear modulus measurements in Co doped BaFe$_2$As$_2$ single crystals \cite{Fernandes-shear,Yoshizawa}. However corresponding measurements of the electronic nematic response function in the charge/orbital or spin degrees of freedom are currently lacking. Here we show that the dynamical charge nematic response can be probed directly using electronic Raman scattering. For both SrFe$_2$As$_2$ (Sr122) and BaFe$_2$As$_2$ (Ba122) systems, in the tetragonal phase of strain-free crystals, the low energy Raman response displays a strongly temperature dependent quasi-elastic peak whose intensity increases upon approaching the structural transition and collapses in the orthorhombic phase. By extracting the static charge nematic susceptibility from our Raman measurements, we are able to unravel the presence of an incipient charge nematic order with very similar characteristic temperature scales for Ba122 and Sr122 systems, indicating the universal character of charge nematic fluctuations in the 122 family of iron-based superconductors.
  
\section{Electronic Raman scattering and charge nematicity}

Electronic Raman scattering in the symmetry channel $\mu$ is proportional to the weighted charge correlation function $S^{\mu}(\omega$)=$\avg{\rho^{\mu}(\omega)\rho^{\mu}(-\omega)}$.
The correlation function $S^{\mu}$ is in turn directly linked to the imaginary part of the Raman response $(\chi^{\mu})''$ via the fluctuation dissipation theorem
\begin{equation}
S^{\mu}(\omega)=-\frac{\hbar}{\pi}(1+n(\omega,T))(\chi^{\mu})''(\omega)\label{bose}
\end{equation}
 where $n(\omega,T)$ is the Bose-Einstein distribution function.
The electronic Raman response function $\chi^{\mu}$ is given by
\begin{equation}
\chi^{\mu}(\omega)=\frac{i}{\hbar}\int_{0}^{\infty}dte^{i\omega t}\left<[\rho^{\mu}(t),\rho^{\mu}(0)]\right>\label{response}
\end{equation}
 where 
$\rho^{\mu}=\sum_{\mathbf{k}}\gamma_{\mathbf{k}}^{\mu}n_{\mathbf{k}}$
with $n_{\mathbf{k}}$ the charge density operator and $\gamma_{\mathbf{k}}^{\mu}({\bf {e_{i},e_{s}})}$ a form factor, also called the Raman vertex. The symmetry index $\mu$ of the form factor depends on the polarizations of the incident and scattered
lights $\bf{e}_{i}$ and $\bf{e}_{s}$ respectively.~\cite{Dev-Hackl} Since
the photon wave vector is several orders of magnitude smaller than
the typical Brillouin zone size, there is negligible momentum transfer
in the electron-photon scatterings, and therefore Raman spectroscopy
probes the system uniformly.

\par
The $x^{2}-y^{2}$  ($B_{1g}$) symmetry can be selected by choosing
crossed incoming and outgoing photon polarizations at $45^{\circ}$ with
respect to the Fe-Fe bonds (see figure \ref{fig1}a). Similarly
the $xy$  ($B_{2g}$) symmetry can be selected by choosing incoming
and outgoing photon polarizations along the Fe-Fe bonds. 
The momentum space structure of $\gamma_{\mathbf{k}}^{\mu}$ is constrained
by symmetry. In the case of $x^{2}-y^{2}$ symmetry, $\gamma_{\mathbf{k}}^{\mu}$
must change sign under mirror symmetry with respect to the direction
at $45^{\circ}$ of the $x$ and $y$ axis. 
The k-space structure of these form factors are shown in figure \ref{fig1}a. In $x^{2}-y^{2}$ symmetry, the effective charge $\rho^{x^2-y^2}$ is the product of the charge density $n_k$  and a form factor which changes sign under a $90^{\circ}$ rotation. This corresponds to the order  parameter of a charge nematic $\phi=\avg{\sum_k \gamma_{\mathbf{k}}^{x^{2}-y^{2}}n_{\mathbf{k}}}$ with orientational order along the Fe-Fe bonds  \cite{Vojta,Fradkin}. Thus in this symmetry electronic Raman scattering is a direct probe of charge nematic order parameter fluctuations \cite{Yamase2013}. Besides,  since $\chi^{\mu}$ is proportional to the square of the charge nematic order parameter $\phi^{2}$, it is possible to extract charge nematic fluctuations directly from the Raman response without applying any external symmetry breaking field such as uniaxial stress. 

\section{Sample and experiments}
Single crystals of BaFe$_2$As$_2$  (Ba122) and Sr(Fe$_{1-x}$Co$_x$)$_2$As$_2$ (Co-Sr122) were grown using the self-flux method \cite{Rullier}. The as-grown BaFe$_2$As$_2$ crystal was further annealed for three weeks at 700$^{\circ}C$ and then slowly cooled down to 300$^{\circ}C$, resulting in an increase of $T_S$ by several degrees. Electronic Raman scattering results on two different single crystals, one for each family, are presented here: undoped BaFe$_2$As$_2$ and Sr(Fe$_{1-x}$Co$_x$)$_2$As$_2$ with x=0.04. Their structural transition temperatures, as determined by transport measurements, are identical: $T_S$=138$\:\mathrm{K}$. Residual resistivity ratios (RRR=$\rho_{300K}/\rho_{0K}$) for the two single crystals indicate different degrees of disorder: RRR $\sim$ 9 for annealed Ba122 and RRR $\sim$ 1 for Co-Sr122. The full Co doping dependence of the electronic Raman response for the Ba122 system can be found elsewhere \cite{Gallais}. In order to extract the imaginary part of the Raman response function, the raw spectra were corrected for the Bose factor and the instrumental spectral response. Data were taken in two different polarization configurations probing respectively $x^2$-$y^2$ ($B_{1g}$) and $xy$ ($B_{2g}$) symmetry. As explained above these two configurations are sensitive to charge nematic fluctuation along Fe-Fe bonds and at $45^{\circ}$ of the Fe-Fe bonds respectively.

\section{Results and discussion}

\begin{figure}
\begin{center}
\epsfig{file=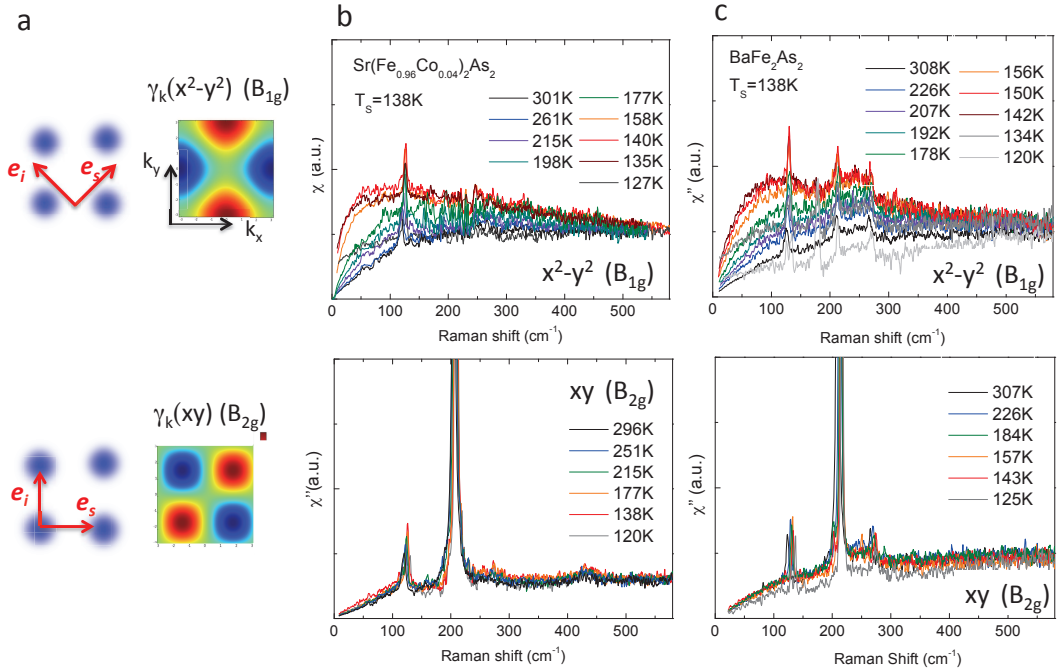,width=14cm}
\end{center}
\caption{(Color online) a: Polarization configurations showing the incoming, $e_i$, and outgoing $e_s$ photon polarizations with respect to the Fe square plane for  $x^2$-$y^2$ ($B_{1g}$) and $xy$ ($B_{2g}$) symmetries. The notation $B_{1g}$ refers to the one Fe unit cell whose axes are along the Fe-Fe bonds. The k-space structure of the Raman form factors (or vertex) $\gamma_k$ is depicted for each symmetry. b: Temperature dependent Raman spectra for Sr(Fe$_{1-x}$Co$_x$)$_2$As$_2$  (x=0.04) and c for BaFe$_2$As$_2$ (data taken from \cite{Gallais}) in the two symmetries.}
\label{fig1}
\end{figure}
Figure \ref{fig1}b, c display low energy temperature dependent Raman spectra for Co-Sr122 in $x^2$-$y^2$ (B$_{1g}$) (b) and $xy$ (B$_{2g}$) (c) symmetries. The data for Ba122 are shown for comparison \cite{Gallais}. In both symmetries, the spectra show a broad electronic continuum with sharp phonon lines superimposed to it. The structural transition induces a splitting of the in-plane Fe-As phonon mode at $\sim$ 130$\:\mathrm{cm^{-1}}$  and an activation of the As out-of plane mode in the $x^2$-$y^2$ symmetry at $\sim$ 180$\:\mathrm{cm^{-1}}$. These changes are particularly visible in the spectrum of Ba122 at 120$\:\mathrm{K}$ and are discussed elsewhere \cite{Chauviere2009,Chauviere2011}. Focusing on the electronic continuum, Fig. \ref{fig1}c shows that the spectra in $xy$ symmetry display an essentially temperature independent and flat continuum in the tetragonal phase $T\geq$$T_S$. By contrast the spectra in $x^2-y^2$ symmetry are strongly temperature dependent: upon cooling the electronic Raman continuum shows a significant built-up of spectral weight at low energy, below $\sim$500$\:\mathrm{cm^{-1}}$. This quasi-elastic peak (QEP) emerges on top of a flat continuum similar to the one observed in $xy$ symmetry, and extends up to much higher energies. Similarly to Ba122, the QEP intensity increases upon approaching $T_S$ before collapsing in the orthorhombic phase. The distinctive symmetry of this QEP and its temperature dependence clearly link it to growing dynamical charge nematic fluctuations of $x^2$-$y^2$ symmetry in the tetragonal phase.
\begin{figure}
\begin{center}
\epsfig{file=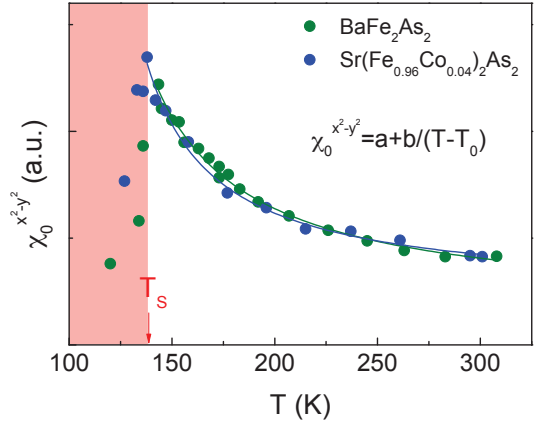,width=7cm}
\end{center}
\caption{(Color online). Temperature dependence of the static charge nematic susceptibility $\chi_{0}^{x^2-y^2}$ for BaFe$_2$As$_2$ and Sr(Fe$_{0.96}$Co$_{0.04}$)$_2$As$_2$. The lines are Curie-Weiss law fits of the susceptibilities in the tetragonal phase (T$\geq$$T_S$).}
\label{fig2}
\end{figure}
\par
A quantitative analysis of the charge nematic fluctuations can be obtained by extracting the static charge nematic susceptibility, $\chi_{0}^{x^2-y^2}$, from $\chi''$ using Kramers-Kronig relation:
\begin{equation}
\chi_{0}^{x^2-y^2}=\frac{2}{\pi}\int_{0}^{\infty} d\omega(\chi^{x^2-y^2})''(\omega)/\omega
\label{KK}
\end{equation}
Because we only have access to a limited energy window, the integral was restricted to energies below 500$\:\mathrm{cm^{-1}}$ since above this energy the spectra are temperature independent in the tetragonal phase. At low energy, below $\sim$10$\:\mathrm{cm^{-1}}$, the spectra were extrapolated using an overdamped Lorentzian form (see below). The resulting static susceptibilities for Ba122 and Co-Sr122 as a function of temperature are compared in Fig.\ref{fig2}. The enhancements of $\chi_{0}^{x^2-y^2}$ are remarkably similar and display a maximum at $T_S$=138$\:\mathrm{K}$. In the tetragonal phase, above $T_S$, the nematic susceptibility can be fitted using: \begin{equation}
\chi_{0}^{x^2-y^2}=a+\frac{b}{T-T_0}
\end{equation}
where $a$ represents the temperature independent flat continuum and the Curie-Weiss term describe the diverging behavior of the QEP. The extracted $T_0$ are very similar, $T_0$=105$\:\mathrm{K}$ ($\pm$ 5$\:\mathrm{K}$) for Co-Sr122 and $T_0$=95$\:\mathrm{K}$ ($\pm$ 5$\:\mathrm{K}$) for Ba122. Comparison of the charge nematic susceptibilities of Ba122 and Co-Sr122 thus indicates an universal behavior among members of the 122 family, and a relative insensitivity of charge nematic fluctuations to static disorder. 
\par
\begin{figure}
\begin{center}
\epsfig{file=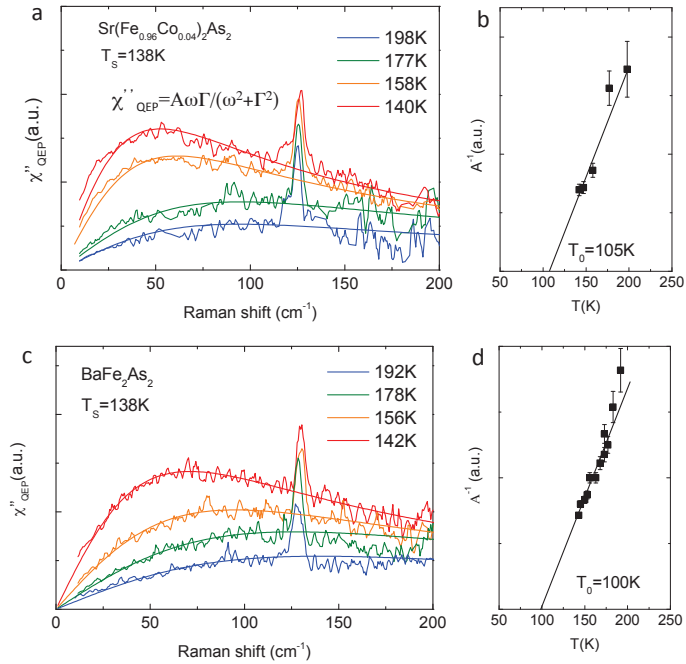,width=9cm}
\end{center}
\caption{(Color online). a,c: Temperature dependence of the extracted quasi-elastic peak (QEP) contribution for BaFe$_2$As$_2$ and Sr(Fe$_{0.96}$Co$_{0.04}$)$_2$As$_2$ along with fits using a relaxational Lorentzian form (see text). b, d: temperature dependences of the inverse of the divergent contribution to the static charge nematic susceptibility, $A^{-1}$, extracted from the Lorentzian fits. The lines are linear (Curie-Weiss) extrapolation of the data between 200$\:\mathrm{K}$ and $T_S$=138$\:\mathrm{K}$, allowing a determination of the characteristic temperature $T_0$ for both systems.}
\label{fig3}
\end{figure}

The divergent part of the charge nematic susceptibility can also be extracted by subtracting the flat temperature independent continuum from the electronic Raman response. This can be done by fitting the electronic Raman continuum at 300$\:\mathrm{K}$, far away from $T_S$ where it becomes almost temperature independent, and subtracting it from the data at  temperatures close to $T_S$. The resulting  QEP contribution is shown in Fig. \ref{fig3} as a function of temperature for Ba122 and Co-Sr122. The QEP can be modeled by a Lorentzian relaxational form with an amplitude $A$ and a scattering rate $\Gamma$:  $\chi_{QEP}''=A \frac{\omega\Gamma}{\Gamma^{2}+\omega^{2}}$.
Within this framework the divergent part of the static charge nematic susceptibility, $\chi_{0, QEP}^{x^2-y^2}$, is given by the amplitude $A$ of the QEP: $\chi_{0, QEP}^{x^2-y^2}$=$\frac{2}{\pi}\int_{0}^{\infty}d\omega\chi''_{QEP}(\omega)/\omega=A$. 
As shown in figure \ref{fig3}, this relaxational form provides a good description of the energy dependence of the QEP for temperatures up to at least $\sim$ 60$\:\mathrm{K}$ above $T_S$. In the case Ba122 the extracted $A^{-1}$, equivalent to the charge nematic stiffness, shows a linear temperature dependence, $A^{-1}\sim$ $T-T_0$, in agreement with a Curie-Weiss-like behavior. We note that deviations from linearity above 180$\:\mathrm{K}$ might be ascribed to the fact that the subtracted background is approximated by the 300$\:\mathrm{K}$ spectrum. This approximation makes departure from Curie-Weiss behavior unavoidable at high temperatures. On the other hand no solid conclusions can be drawn from the behavior of $A^{-1}$ for Co-Sr122 due to a more limited number of data points. Nevertheless for both Ba122 and Co-Sr122 the linear interpolation of $A^{-1}$ yields similar values of $T_0$ as the ones found in the previous analysis: 100$\:\mathrm{K}$($\pm$ 5$\:\mathrm{K}$) and 105$\:\mathrm{K}$ ($\pm$ 10$\:\mathrm{K}$) respectively.
\par
It is interesting to note that shear modulus measurements on Ba122 found a similar Curie-Weiss-like softening \cite{Fernandes-shear,Yoshizawa}. However the Curie-Weiss temperature extracted from shear modulus measurements was found to be very close to $T_S$. This is in contrast to the one extracted here which is well below $T_S$ by at least $\sim$ 40$\:\mathrm{K}$, implying that the charge nematic fluctuations grow slower than the lattice ones \cite{Gallais}. It indicates that the observed charge nematic fluctuations are not sufficient to drive the structural transition, and that the charge nematic order which would otherwise occurs at a lower temperature $T_0$, is pre-empted by the structural transition. It therefore seems that another degree of freedom is responsible for the softening of the shear modulus. A recent analysis of NMR spin-lattice relaxation rate data indicates that spin fluctuations can account for the observed lattice softening via magneto-elastic coupling \cite{Fernandes-Meingast, Ning, Nakai, Paul}. It remains to be seen if magnetic fluctuations can also account for the observed charge nematic softening via spin-charge coupling. 
\par
In summary, electronic Raman scattering experiments unveil the presence of charge nematic fluctuations in the tetragonal phase of Ba122 and Sr122 systems. Their divergent behavior indicates the presence of an incipient charge nematic instability in the phase diagram of 122 iron superconductors. Our results also establish Raman scattering as a key probe of uniform charge order and fluctuations such as nematics, which might be relevant in other strongly correlated electron systems.
\par
We thank F. Rullier-Albenque for providing us with transport data. Y.X. Y., Y.G., L.C., M.C, M.-A.M., A.S., D.C. and A.F. acknowledge support from Agence Nationale de la Recherche through ANR Grant PNICTIDES.

\end{document}